\documentclass[sigconf]{acmart}

\usepackage{graphicx}
\usepackage{amsmath}
\usepackage{booktabs}
\usepackage{algorithm}
\usepackage{algorithmic}
\usepackage{amsfonts}
\usepackage{multirow}
\usepackage{makecell}
\usepackage{subfigure}
\usepackage{color}
\usepackage{bm}
\usepackage{epstopdf}
\usepackage{url}
\usepackage[cal=cm]{mathalfa}
\usepackage{balance}
\usepackage{threeparttable}
\usepackage{lipsum}
\usepackage{enumitem}
\usepackage{xcolor}       
\usepackage{tcolorbox}    
\tcbuselibrary{listings} 
\usepackage{listings}    

\tcbuselibrary{most}
\definecolor{skyblue}{RGB}{64,158,255}
\definecolor{lightsky}{RGB}{230,245,255}

\newtcolorbox{codebox}[1][]{
    enhanced,
    breakable,
    colback=white, 
    colframe=gray!20, 
    colbacktitle=skyblue, 
    coltitle=white, 
    fonttitle=\bfseries\sffamily,
    fontupper=\ttfamily, 
    sharp corners, 
    boxrule=0.5pt, 
    title=#1, 
    attach boxed title to top left={ 
        xshift=5mm, 
        yshift*=-3mm
    },
    boxed title style={ 
        sharp corners, 
        colframe=skyblue!80!black,
        colback=skyblue,
        boxrule=0.5pt,
        left=3mm, 
        underlay={ 
            \draw[fill=skyblue!80!black, draw=none] 
                (frame.south west) rectangle 
                ([xshift=3mm]frame.north west);
        }
    },
    underlay={ 
        \draw[fill=skyblue!10, draw=none] 
            (frame.south west) rectangle 
            ([xshift=3mm]frame.north west);
    },
    drop fuzzy shadow, 
    top=8pt,
    left=3mm, 
    right=6pt,
    before skip=12pt,
    after skip=12pt
}

\author{Shuo Yang}
\affiliation{
  \institution{Nanjing University of Science and Technology}
  \city{Nanjing}
  \country{China}
 }
\email{yangshuo1115@gmail.com}

\author{Jiangxia Cao$^\star$}
\thanks{$^\star$Corresponding authors.}
\affiliation{
  \institution{Kuaishou Technology}
  \city{Beijing}
  \country{China}
 }
\email{jiangxiacao@gmail.com}

\author{Haipeng Li}
\affiliation{
  \institution{Nanjing University of Science and Technology}
  \city{Nanjing}
  \country{China}
 }
\email{lihaipeng@njust.edu.cn}

\author{Yuqi Mao}
\affiliation{
  \institution{Nanjing University of Science and Technology}
  \city{Nanjing}
  \country{China}
 }
\email{yuqimao@njust.edu.cn}

\author{Shuchao Pang$^\star$}
\affiliation{
  \institution{Nanjing University of Science and Technology}
  \city{Nanjing}
  \country{China}
 }
\email{pangshuchao@njust.edu.cn}

\usepackage{colortbl}

\definecolor{gblue}{RGB}{66,133,244}

\usepackage{arydshln}
\makeatletter
\def\adl@drawiv#1#2#3{%
        \hskip.5\tabcolsep
        \xleaders#3{#2.5\@tempdimb #1{1}#2.5\@tempdimb}%
                #2\z@ plus1fil minus1fil\relax
        \hskip.5\tabcolsep}
\newcommand{\cdashlinelr}[1]{%
  \noalign{\vskip\aboverulesep
           \global\let\@dashdrawstore\adl@draw
           \global\let\adl@draw\adl@drawiv}
  \cdashline{#1}
  \noalign{\global\let\adl@draw\@dashdrawstore
           \vskip\belowrulesep}}
\makeatother

\usepackage{xcolor}
\newcommand{\myred}[1]{\textcolor{red}{#1}}

\AtBeginDocument{%
  \providecommand\BibTeX{{%
    \normalfont B\kern-0.5em{\scshape \kern-0.25em b}\kern-0.8em\TeX}}}
    
\begin{document}
\title{RecCoT: Enhancing Recommendation via Chain-of-Thought}
\renewcommand{\shorttitle}{RecCoT}

\begin{abstract}
In real-world applications, users always interact with items in multiple aspects, such as through implicit binary feedback (e.g., clicks, dislikes, long views) and explicit feedback (e.g., comments, reviews).
Modern recommendation systems (RecSys) learn user-item collaborative signals from these implicit feedback signals as a large-scale binary data-streaming, subsequently recommending other highly similar items based on users’ personalized historical interactions.
However, from this collaborative-connection perspective, the RecSys does not focus on the actual content of the items themselves but instead prioritizes higher-probability signals of behavioral co-occurrence among items.
Consequently, under this binary learning paradigm, the RecSys struggles to understand why a user likes or dislikes certain items.
To alleviate it, some works attempt to utilize the content-based reviews to capture the semantic knowledge to enhance recommender models.
However, most of these methods focus on predicting the ratings of reviews, but do not provide a human-understandable explanation.
To identify users' reviews better, we explore a cascading reasoning framework, RecCoT, to generate a human-interpretable Chain-of-Thought to enhance recommendations.
Inspired by the great success of reasoning LLMs (e.g., DeepSeek R1), we first propose a frequency-aware reward policy to supervise a larger LLM to adapt the long-tail rating distribution.
After the larger LLM convergenced, we could generate the explainable reasoning CoT for each review, to support fine-tuning a smaller LLM to produce the semantic embeddings to enhance recommendation accuracy.
Our work investigates the impact of reasoning ability on personalized recommendations, achieving significant improvements in recommendation performance 
\footnote{\url{https://github.com/shuoyang2/RecCoT}}.
%
%
\end{abstract}


\maketitle

\section{Introduction}

%
Nowadays, with the booming of internet technology and digital content growth, the information explosion is becoming a more and more serious problem affecting user experience.
To help users find the most relevant information, the RecSys became more and more important to many platforms, such as Kuaishou (short-video and live-streaming), Taobao (shopping)  and so on.
To our knowledge, in industry, these platform RecSys are mostly based on collaborative filtering idea(e.g.,YouTube~\cite{youtube}), which learns user preferences through multiple-binary classification tasks of massive user behaviors, (e.g., is click/dislike/long-view/buy the item?).
In general, CF is extremely simple, efficient, and accurate in capturing users’ positive behavior signals.
However, it primarily focuses on probability calculations rather than truly understanding the meaning behind the items, making it difficult to interpret negative behaviors (e.g., dislikes).
Particularly, instead of implicit binary feedback (e.g., clicks, dislikes, long views), users also leave some explicit feedback (e.g., comments, reviews) to describes his/her feelings about this interaction further, which leave space to understand why a user likes or dislikes certain items.
Hence, a hot topic is to utilize the content-based reviews to provide semantic knowledge to enhance RecSys.

%
In this research direction, a major task is Rating prediction, which is formulated as follows: given a user’s historical reviews and item reviews from other users, predict the rating for a user-item pair.
Early works such as DeepCoNN~\cite{DeepCoNN} and D-attn~\cite{dattn} follow a disentangled two-tower paradigm. 
They generate user and item representations separately and compute the final score via dot product (DeepCoNN and D-attn employ CNNs and Transformers~\cite{attention} as their towers, respectively).
In this paradigm, ANR~\cite{anr} extends D-attn by generating multiple representations for users/items and using user/item-aware weights to fuse them.
The above methods rely on small models trained from scratch. However, with the rise of large language models (LLMs), recent works tend to leverage pre-trained LLMs to enhance model capabilities, as semantic knowledge provides generalized information.
For example, LLM-based approaches like RGCL~\cite{RGCL} use BERT~\cite{bert} to encode reviews, while LoID~\cite{LoID} employs LoRA~\cite{LoRA} to fine-tune the LLM, delivering more powerful semantic representations.
%


Although these methods raised model ability with content information, they ignore the following problems: 
\begin{itemize}[leftmargin=*,align=left]
    \item \textit{Unexplainable Semantic Compression}:
    Current LLM-based RecSys suffer from opaque decision pathways: they compress text semantics into unexplainable latent representations, obscuring the logical connections in the prediction process.
    \item \textit{Highly Computation Cost}:
    Previous works always leverage large language models (LLMs) in an end-to-end manner incur exorbitant computational demands, primarily due to the immense scale of billion-parameter models(e.g., GPT~\cite{gpt-4}), 
    leading to prohibitively long training times and high inference latency, making them impractical for many real-world scenarios\footnote{Training LoID for one epoch on 100M dataset requires 24h with a NVIDIA T4 GPU.}. 
\end{itemize}
In this paper, we present RecCoT, a novel recommendation framework featuring two key technical contributions:
\begin{enumerate}[leftmargin=*,align=left]
\item \textit{Slow Interpretable CoT Generation via Larger LLM}:
Inspired by the great success of reasoning ability in NLP, we consider to utilize such training paradigm to enhance recommendation, e.g., GRPO~\cite{deepseekmath}.
However, in RecSys, there exists a serious long-tail problem (e.g, rating=5 data are more than 90\% in many datasets), thus naive GRPO training strategy suffers from reward hacking~\cite{think} problem.
To this end, we first propose a frequency-aware reward policy to supervise a larger LLM to adapt the long-tail rating distribution.
It introduces an accuracy reward mechanism to ensure explainable reasoning generation.
%
%
\item \textit{Fast Semantic Embedding Alignment via Smaller LLM}:
Since the reasoning model always require billion-scale model parameter, thus the end-to-end training paradigm is impossible for the downstreaming RecSys.
Therefore, in this paper, we apply a cache paradigm to provide the semantic knowledge for the downstreaming RecSys to avoid the massive computation.
To ensure the generated semantic embedding could align with the downstreaming task, we further fine-tune a smaller LLM to generate the semantic embeddings for each reviews.
Thanks for our Slow-Fast LLM designing, our RecSys model can be greatly streamlined to achieve lightweight and efficient prediction.
\end{enumerate}
\textbf{Contributions.} 
Our contributions are summarized as follows:
\begin{itemize}[leftmargin=*,align=left]
\item We propose a decoupled reasoning-execution framework: Larger LLMs handle interpretable CoT generation while smaller LLMs perform high-speed semantic alignment. 
\item We devise a reward strategy to adapt long-tail distribution that ensures the quality of CoT. 
\item We conduct detailed analyses with SOTA methods.

\end{itemize}

\section{RELATED WORK}
\subsection{LLM for RecSys.}
The field of recommendation systems has recently seen a significant shift with the advent of Large Language Models (LLMs).
While traditional methods primarily rely on collaborative filtering, matrix factorization, and deep learning architectures like GRUs and LSTMs, LLMs offer a new paradigm for modeling user-item interactions~\cite{P5,M6-Rec}.
A key application is using LLMs as powerful feature extractors, where they are pre-trained on vast text corpora to generate rich, context-aware representations from textual data such as product descriptions and user reviews~\cite{RecLLMs}.
This approach enriches item and user profiles, addressing sparsity issues and capturing nuanced semantic information.

Furthermore, LLMs are increasingly being used for sequential recommendation tasks, leveraging their strong capabilities in modeling sequential data to predict future user actions based on their past behavior\cite{hu2024enhancing}.
Another emerging direction is the use of LLMs as a generative recommender, where the model directly generates a list of recommended items or provides personalized suggestions based on a user-centric prompt.
This allows for more conversational and dynamic recommendation experiences. 
Recent work has also explored using LLMs to introduce external knowledge and enhance the reasoning abilities of recommender models, moving beyond simple pattern matching to more knowledge-driven recommendations\cite{RecSAVER}.

\subsection{Chain-of-Thought (CoT)}
Chain-of-Thought (CoT)~\cite{CoT} is a prompting technique that has been highly successful in improving the reasoning capabilities of LLMs by encouraging them to break down complex problems into a series of intermediate steps. 
Originally demonstrated in tasks requiring multi-step reasoning~\cite{dua2022successive}, such as arithmetic, commonsense reasoning, and logical deduction, CoT has proven to be an effective strategy for eliciting more accurate and robust responses from large models. 
By generating these intermediate reasoning steps, the model not only arrives at a better answer but also provides a transparent, human-readable trace of its thought process.

While CoT's primary application has been in general-purpose NLP tasks, its principles are highly relevant to recommendation systems.
Generating a CoT for a recommendation can provide a crucial link between a user's historical data and a final prediction~\cite{RecSAVER}. 
For instance, CoT could explicitly reason about a user's past review sentiments or item preferences to justify a recommendation, thereby directly addressing the "black-box" problem that often plagues modern recommender models. 
This integration of CoT with LLMs for RecSys is a promising direction, offering a path toward more interpretable, explainable, and trustworthy recommendation systems.

\section{Preliminary}
\subsection{Problem Statement}
In this work, we consider a rating prediction task as the primary research goal.
Specifically, given \textbf{observed} \myred{Review-Rating ($t_{u,i}$, $r_{u,i}$)} of \myred{user-item pair ($u$,$i$)} as corpus to train a model, our goal is to predict other \textbf{un-observed} \myred{user-item rating score $r_{\mathrm{u,i}}$} according to user historical reviews and item reviews from other users.
This task presents a significant challenge due to the inherent sparsity of user-item interactions in real-world datasets. 
To overcome this, our model learns to predict ratings by analyzing both the user's historical reviews and the reviews provided by other users for a specific item.
This approach allows the model to deeply understand user preferences and item characteristics from unstructured text data, rather than relying solely on past ratings. 
The model must therefore effectively fuse these disparate sources of information to make accurate and meaningful predictions for items a user has never seen before.

\subsection{GRPO}
\begin{figure}[t!]
\begin{center}
\includegraphics[width=8cm]{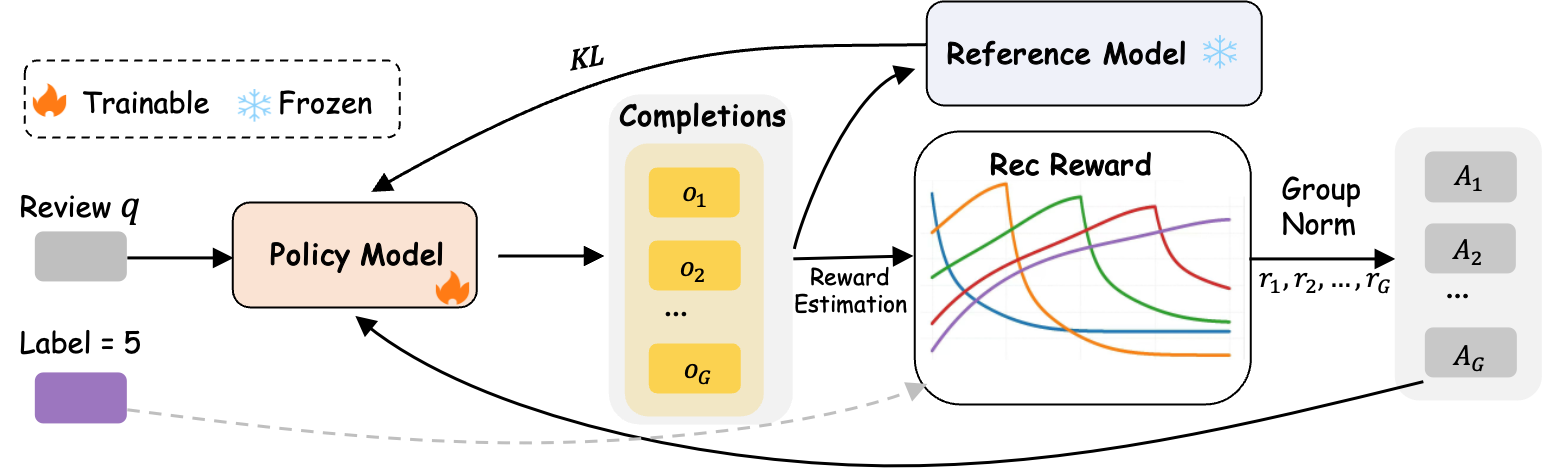}
\caption{Overall framework of GRPO used in RecCoT.}
\label{grpo}
\end{center}
\vspace{-0.5cm}
\end{figure}
GRPO is a reinforcement learning method that directly uses the average reward of multiple sampling outputs of the same group as the baseline, which significantly reduces the use of training resources and simplifies the training process.
Specifically, for each \myred{problem $q$}, GRPO samples a set of \myred{CoTs $\mathbf{\left\{o_{1}, o_{2}, \cdots, o_{G}\right\}}$} from the \myred{old policy $\mathbf{{\pi}_{{\theta}_{old}}}$}, and then optimizes the policy model by maximizing the following objective function:
%
%
\begin{equation} 
\footnotesize
\begin{aligned}
&\mathcal{J}_{G R P O}(\theta)=
\min \Bigl(\frac{\pi_{\theta}\left(o_{i} \mid q\right)}{\pi_{\theta_{o l d}}\left(o_{i} \mid q\right)} A_{i}, \operatorname{clip}\Bigl(\frac{\pi_{\theta}\left(o_{i} 
\mid q\right)}{\pi_{\theta_{o l d}}\left(o_{i} \mid q\right)}, 1-\varepsilon, 1+\varepsilon\Bigr) A_{i}\Bigr)\\
&\quad \quad \quad \quad \quad \quad- \beta \mathbb{D}_{K L}\left(\pi_{\theta} \| \pi_{r e f}\right),
\end{aligned}
\label{eqgrpo1}
\end{equation}
\begin{equation}
\footnotesize
\mathbb{D}_{K L}\left(\pi_{\theta} \| \pi_{r e f}\right)=\frac{\pi_{r e f}\left(o_{i} \mid q\right)}{\pi_{\theta}\left(o_{i} \mid q\right)}-\log \frac{\pi_{r e f}\left(o_{i} \mid q\right)}{\pi_{\theta}\left(o_{i} \mid q\right)}-1,
\label{eqgrpo2}
\end{equation}
\begin{equation}
\footnotesize
A_{i}=\frac{r_{i}-\operatorname{mean}\left(\left\{r_{1}, r_{2}, \cdots, r_{G}\right\}\right)}{\operatorname{std}\left(\left\{r_{1}, r_{2}, \cdots, r_{G}\right\}\right)}.
\label{eqgrpo3}
\end{equation}
where $q$ is input tokens, $o_{i}$ is the CoT results, $A_{i}$ is the normalized rewards of CoT $o_{i}$ within each group $\{o_{1}\dots,o_{G}\}$.
In this way, the new policy model can increase the probability of higher reward CoT.
The figure~\ref{grpo} illustrates the framework of GRPO in this paper.

\subsection{Prompt}
Our prompting strategy is carefully designed to guide a Large Language Model (LLM) in performing the recommendation task.
The prompt integrates key information, such as the product ID and user review, to direct the model's reasoning.
A core component of this strategy is the use of Chain-of-Thought (CoT), which encourages the LLM to generate a clear rationale for its predicted rating.
This allows the model to output not only a predicted rating but also a transparent, human-readable reason for its decision. 
An abstract prompt template is illustrated in Table~\ref{tab:prompt}.
\begin{table}[t]
\centering
\caption{Rating Prediction Prompt and Response}
\label{tab:prompt}
\begin{tabular}{|l|p{6.5cm}|}
\hline
\textbf{Role} & \textbf{Content} \\
\hline
\multirow{2}{*}{User} & \textbf{Task:} You are an expert in rating prediction. For the product with ID \texttt{\{instruction\}} and user review "\texttt{\{review\}}", analyze the review and predict its potential rating on a scale of 1 to 5. \\
\cline{2-2}
& \textbf{Output Format:} Please output the result in JSON format with the following structure: \texttt{\{\newline\quad"think": "analysis process",\newline\quad"answer": "predicted rating"\newline\}} \\
\hline
Assistant & Okay, I will analyze according to the following steps: \\
\hline
\end{tabular}
\end{table}
\section{Methodology}
\begin{figure*}[t]
\begin{center}
\includegraphics[width=18cm]{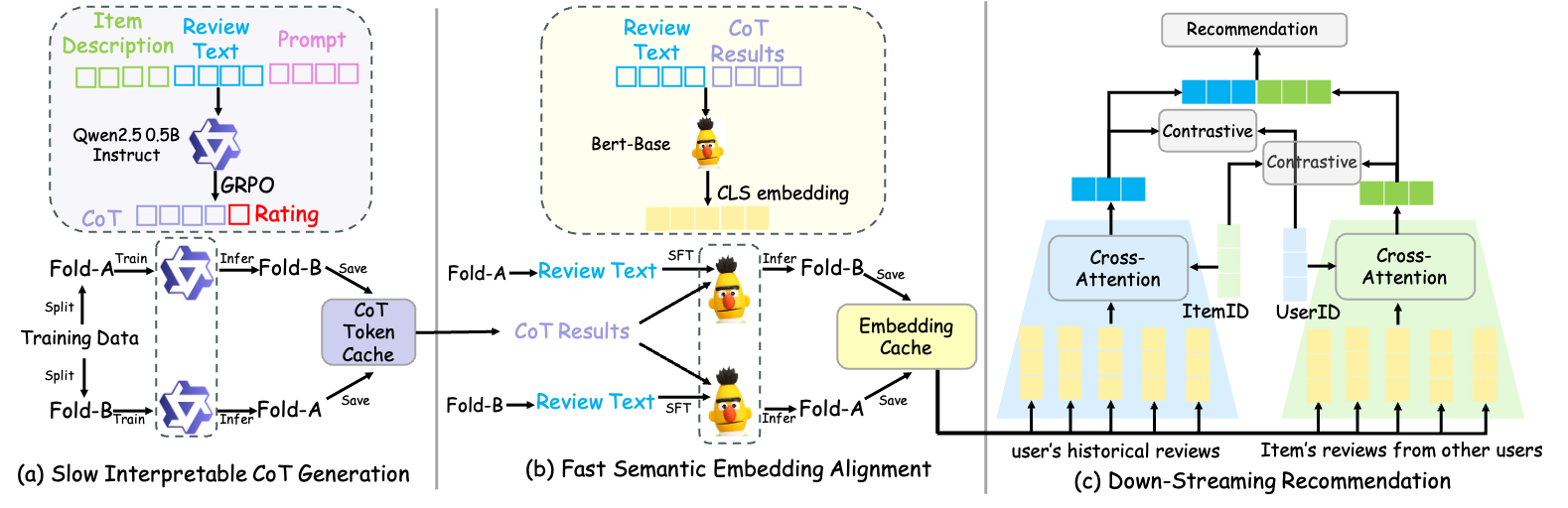}
\caption{Model Architecture of the proposed RecCoT, including three steps: (a) CoT generation via a slow larger LLM, (b) embedding compression via a fast smaller LLM, and (c) down-streaming recommendation.}
\label{method}
\end{center}
\vspace{-0.5cm}
\end{figure*}
The architecture of our method RecCoT is illustrated in Figure~\ref{method}.
In part (a), we utilize LoRA to fine-tune a larger Qwen~\cite{qwen}, adapting it to specific datasets and outputting it in a given format.
In part (b), we combine interpretable reasoning content from part(a) with original reviews as input to fine-tune a smaller BERT. 
The resulting CLS representations are then cached in a storage to support downstreaming RecSys. 
In part (c), these stacked CLS vectors are processed through attention layers and fed into a contrastive learning module to generate final recommendation scores.

\subsection{CoT Generation}
%
A key challenge we face is ensuring the Qwen model generates high-quality Chain-of-Thought (CoT) explanations, particularly for the long-tail distribution that is common in recommender systems.
To address this challenge, we've developed a multi-reward system to guide the CoT generation process. 
This system uses three distinct reward functions and is inspired by the methodology of Deepseek R1~\cite{deepseekr1}.

\textbf{Format Reward.}
We implemented regular expression-based pattern matching to validate whether the output CoT adheres to predefined JSON format.
If compliance with the predefined JSON format, the $R_{\mathrm{format}} \,=\, 1$; otherwise, it is assigned 0.

\textbf{Analysis Quality Reward.}
Quantify the logical depth of CoT through semantic integrity indicators.
\begin{equation}
R_{\text{quality}} = \frac{\ell_{\text{CoT}} - \ell_{\min}}{\ell_{\max} - \ell_{\min}}
\end{equation}
where $\mathbf{\ell_{\text{CoT}}}$ is the length of the current CoT text, $\ell_{\text{min}}=100$
and $\ell_{\text{max}}=200$ are the preset minimum and maximum effective length thresholds, respectively. 
This function ensures that CoT avoids information redundancy and has sufficient analytical depth.

\textbf{Frequency-aware Accuracy Reward.}
To alleviate the long-tail problem, we devised a novel reward strategy that goes beyond simple accuracy. 
This approach encourages the model to provide more nuanced predictions by increasing the reward for low-frequency categories and decreasing it for high-frequency ones. 
We further incentivize the model to output float results for relative value comparisons, rather than discrete integers. 
The total reward is a weighted sum of the following components:
\begin{itemize}
    \item \textbf{Accuracy:} This is the primary reward component, directly measuring the proximity of the model's prediction to the true rating.
    The model is rewarded for making predictions that are close to the actual user rating.

    \item \textbf{Frequency:} To counter the long-tail problem, the reward is scaled based on the frequency of the true label. 
    Accurate predictions for low-frequency categories receive a higher reward, while those for high-frequency categories are given a lower reward, encouraging the model to perform better on less common items.

    \item \textbf{Penalty:} To refine the model's predictions, a penalty term is introduced that specifically discourages significant over-estimation and under-estimation of the rating.
    This asymmetric penalty helps to prevent systematic biases in the model's predictions, leading to more nuanced and reliable results.
\end{itemize}

Specifically, our piecewise reward function is defined as: 
%
\begin{equation}
\begin{split}
R_{\mathrm{predict}} &= {\frac{1}{\sqrt{f}}} \cdot \left[\texttt{Base}(e) + \texttt{Penalty}(e) \right],\quad e = \mathit{\hat{r} - r},\\
\quad \quad \quad & \texttt{Base}(e) = \exp(-e^2), \quad  \\
\texttt{Penalty}(e) &= 
\begin{cases} 
-\lambda \cdot \left( \exp(\mu \cdot |e|) - 1 \right) \quad  &  e < 0  \\
-\gamma \cdot \left( 1 - \frac{\exp(-\kappa \cdot e)}{1 + e} \right) \quad &  e \geq 0 
\end{cases} \\
\end{split}
\end{equation}
where \myred{$f$ is the occurrence frequency of rating category, $r$ is the true rating, $\mathit{\hat{r}}$ is the predicted rating, $\lambda$ and $\mu$ are underestimation penalty factor and $\gamma$ and $\kappa$ are the overestimation penalty factor}.
Our reward curves are showing in Figure~\ref{grpo}.

In this paper, our GRPO model was trained using a combination of standard and reinforcement learning-specific hyperparameters. 
The model was configured with a maximum sequence length of 512, a learning rate of 2e-4, and was trained for 1 epoch. 
A per-device batch size of 8 with gradients accumulated over 4 steps was used, resulting in an effective batch size of 32, and FP16 mixed precision was employed for efficiency. 
For the core GRPO algorithm, we adopted a PPO clip ratio of 0.2 and a KL divergence coefficient of 0.1 to ensure stable policy updates. 
To guarantee reproducibility, a random seed of 42 was set, and logs were recorded every 50 steps. 

Finally, we could utilize them to serve as reward system to conduct the GRPO (in Eq.(\ref{eqgrpo1}-\ref{eqgrpo3})) to finetune LLMs (e.g., Qwen) to generate CoT for each review with \myred{2-fold setting to avoid the over-fitting problem}.

\subsection{Semantic Embedding Alignment}
%
\textbf{BERT SFT}. After obtaining the Chain-of-Thought (CoT) for each review, we perform a supervised fine-tuning stage on the BERT model to fuse the CoT and original review into a single, rich semantic embedding.
This process is crucial for constructing an augmented dataset that embodies the comprehensive reasoning trajectories, while also mitigating the computational overhead on the downstream recommendation model.

Specifically, \myred{\textbf{the CoT, denoted as $o_{u,i}$}}, and the original review text, $t_{u,i}$, are concatenated and fed into BERT. 
The output from \myred{the BERT model's special [CLS] token, $\mathbf{CLS}_{u,i}$}, serves as a compressed semantic embedding, encapsulating both the raw textual information and the deep cognitive insights from the CoT:
%
\begin{equation}
\mathbf{CLS}_{u,i} = \texttt{BERT}(o_{u,i} \oplus t_{u,i})
\label{eq:concat}
\end{equation}
%
%
Afterward, we use the MSE loss to fine-tune the BERT:
 \begin{equation}
\begin{split}
\mathcal{L}_\mathrm{BERT}=(\hat{r}_\mathrm{u,i} - r_\mathrm{u,i})^2,\quad \hat{r}_\mathrm{u,i}=\texttt{MLP}(\mathbf{CLS}_{u,i})
\label{L_pre}
\end{split}
\end{equation}
where \texttt{MLP} is a projector, $\hat{r}_\mathrm{u,i}$ denotes predicted rating.

\textbf{Embedding Cache}. 
%
After the BERT model has converged, we proceed to cache the semantic embeddings for each review. 
This is a crucial pre-computation step designed to store the rich semantic representations generated by our fine-tuned model, thereby significantly reducing the computational overhead for the downstream recommender system. 
Specifically, we use the fine-tuned BERT to obtain the semantic embedding for every review and store them in a \myred{\texttt{Cache\_Store}} for each user and item, as follows: 
\begin{equation}
\begin{split}
\texttt{Cache\_Store}_u \xleftarrow{}(\mathbf{CLS}_{u}^1, \mathbf{CLS}_{u}^2, \dots, \mathbf{CLS}_{u}^m) \\
\texttt{Cache\_Store}_i \xleftarrow{}(\mathbf{CLS}_{i}^1, \mathbf{CLS}_{i}^2, \dots, \mathbf{CLS}_{i}^n)
\end{split}
\label{cls}
\end{equation}
where the \texttt{Cache\_Store} saves all reviews embedding, the $\mathbf{CLS}_{u}^*$/$\mathbf{CLS}_{i}^*$ are the semantic embeddings of corresponding user/item, and $m$, $n$ represent the number of reviews from user $u$ and item $i$.

\subsection{Down-streaming RecSys with Cache Embedding}
On top of the \texttt{Cache\_Store}, our RecSys model directly retrieves the pre-computed semantic embeddings, eliminating the need for re-computation during training and inference. 
This process, illustrated below, allows us to efficiently access the rich semantic information from user and item reviews.
\begin{equation}
\begin{split}
\mathbf{H}_u &= (\mathbf{CACHE}_u^1, \ldots, \mathbf{CACHE}_u^m) \xleftarrow{} \texttt{Cache\_Store}_u \\
\mathbf{H}_i &= (\mathbf{CACHE}_i^1, \ldots, \mathbf{CACHE}_i^n)  \xleftarrow{} \texttt{Cache\_Store}_i
\end{split}
\label{item_stack}
\end{equation}
Here, $\mathbf{H}_u\in \mathbb{R}^{m\times d}$ and $\mathbf{H}_i \in \mathbb{R}^{n\times d}$ are the semantic embedding matrices for user $u$ and item $i$ respectively, where $d=768$ is the embedding dimension. $m$ and $n$ are the number of reviews for the user and item.

To effectively model user-item relationships, we use cross-attention layers to integrate this rich semantic information into the user and item representations. 
Specifically, we use \myred{the user’s review embedding sequence $\mathbf{H}_u$} to refine \myred{the item ID representation $\mathbf{p}_{i}$}, and vice versa for \myred{the user ID representation $\mathbf{p}_{u}$} with \myred{the item's review embedding sequence $\mathbf{H}_i$}.
\begin{equation}
\begin{split}
\mathbf{v}_i = \text{Attention}(\mathbf{p}_i, \mathbf{H}_u, \mathbf{H}_u), \ \ \mathbf{v}_u = \text{Attention}(\mathbf{p}_u, \mathbf{H}_i, \mathbf{H}_i)
\end{split}
\end{equation}
where $v_i$ and $v_u$ are the refined, context-aware item and user representation that now incorporate semantic informations from user reviews.

To further enhance the quality of these representations, we apply a contrastive learning objective.
We apply a contrastive learning objective to model user-item relationships that might be overlooked by the attention mechanism. 
For each updated representation (anchor), a positive pair is its interacted counterpart, and a negative pair is a randomly sampled representation from the same batch.
We minimize \myred{the following hinge-like loss function ($\mathcal{L}\mathrm{cl}$)} to enforce a separation margin between positive and negative pairs.
\begin{equation}
\mathcal{L}\mathrm{cl} = \max\left(0, \Delta + \mathrm{D}(\mathbf{v}, \mathbf{p}^+) - \mathrm{D}(\mathbf{v}, \mathbf{p}^-)\right)
\end{equation}
In this equation, $\mathbf{v}$ represents the anchor embedding, $\mathbf{p}^+$ and $\mathbf{p}^-$ are the positive and negative embeddings respectively, $\mathrm{D}(\cdot, \cdot)$ is the squared Euclidean distance $||\cdot||_2^2$, and $\Delta$ is a margin hyperparameter that controls the desired separation.

\subsection{Model Optimization}
We employ a MSE loss for the downstream rating prediction task. 
The model predicts \myred{a rating $\hat{r}_{\mathrm{u,i}}$} by passing the concatenated user and item representations through a Multi-Layer Perceptron (MLP). 
The MSE loss then minimizes the squared difference between the predicted and actual ratings.
\begin{equation}
\begin{split}
\hat{r}_{\mathrm{u,i}} = \texttt{MLP}(\mathbf{v}_u \oplus \mathbf{v}_i), \ \
\mathcal{L}_{\mathrm{rec}} = (\hat{r}_{\mathrm{u,i}} - r_{\mathrm{u,i}})^2,
\end{split}
\label{L_rec}
\end{equation}
The overall training objective is to minimize \myred{the total loss, $\mathcal{L}$}, which is a weighted sum of the recommendation loss and the contrastive loss. 
This joint objective ensures that the model is optimized for both predictive accuracy and the quality of its learned representations.
The total loss is defined as:
\begin{equation}
\mathcal{L} = \mathcal{L}_{\mathrm{rec}} + \lambda \mathcal{L}_{\mathrm{cl}}
\label{L_total}
\end{equation}
where $\mathcal{L}_{\mathrm{rec}}$ denotes the prediction loss, $\lambda$ is the weight assigned to the contrastive loss, and $\mathcal{L}_{\mathrm{cl}}$ represents the contrastive loss.

\section{Experiments}
%

\subsection{Datasets}
Our experiments are conducted on the widely recognized Amazon item review dataset, a standard benchmark in the field of recommender systems. 
This dataset provides a rich source of user feedback, encompassing both explicit ratings and textual reviews.
Furthermore, it includes extensive item metadata, such as descriptions, category information, price, and brand details. 
The primary focus of our experimental evaluation is the rating prediction task. 
To ensure the generalizability and representativeness of our findings, we selected seven distinct and representative item categories from the full Amazon dataset~\cite{amazon} for our study.
The specific statistics for these categories are summarized in Table~\ref{tab:example}.

\subsubsection{Data Preprocessing}

\begin{table}[htbp]
\centering
\caption{Statistics of the processed datasets.}
\resizebox{8.0cm}{!}{\begin{tabular}{lccc}
\hline
\textbf{Datasets} & \textbf{U} & \textbf{V} &  \textbf{$r_{u,i}$} \\
\hline
Amazon\&Fashion & 15520 & 86316 & 115464  \\ 
Baby\&Products & 184851 & 168134 & 1551060  \\ 
CDs\&and\&Vinyl & 183733 & 534796 & 2518472 \\ 
Industrial\&and\&Scientific & 105305 & 239885 &889120 \\ 
Musical\&Instruments & 84574 & 149299 & 792162 \\ 
Software & 160256 & 60177 & 1421148\\ 
Video\&Games & 117742 & 100818 & 1030529 \\ \hline
\end{tabular}}
\label{tab:example}
\end{table}

To mitigate data sparsity and enhance dataset quality, we applied a filtering criterion: we removed all users and items with fewer than five interactions.
Additionally, we preprocessed the review text by removing irrelevant characters and very short reviews to improve the quality of textual features. 
\begin{table*}[!t]
\centering
\caption{The experimental results on Amazon datasets (MSE, lower is better.)}
\renewcommand{\arraystretch}{1.3}
\resizebox{\linewidth}{!}{
{
\begin{tabular}{cccc|cc|cc|ccc}
\toprule 
\multirow{2}{*}{Datasets} & \multirow{2}{*}{ANR} & \multirow{2}{*}{BiGI} & \multirow{2}{*}{RGCL} & \multicolumn{2}{c|}{LoID} & \multicolumn{2}{c|}{LoID + Cache Embedding} & \multicolumn{3}{c}{RecCoT} \\
& & & &Emb &RecSys &Emb &RecSys &CoT(MAE) &Emb &RecSys \\
\bottomrule
\cellcolor{brown!15}Amazon\&Fashion & \cellcolor{brown!15}1.68 & \cellcolor{brown!15}1.48 & \cellcolor{brown!15}1.44 
&0.418 & \cellcolor{brown!15}0.680 
&0.418 & \cellcolor{brown!15}\underline{0.633}
&0.800 &0.401 & \cellcolor{brown!15}\textbf{0.570} \\

\cellcolor{green!15}Baby\&Products & \cellcolor{green!15}1.52 & \cellcolor{green!15}1.36 & \cellcolor{green!15}1.38 
&0.272 &\cellcolor{green!15}1.010
&0.272& \cellcolor{green!15}\underline{0.946}
&0.860 &0.268 & \cellcolor{green!15}\textbf{0.941} \\

\cellcolor{yellow!15}CDs\&and\&Vinyl & \cellcolor{yellow!15}0.94 & \cellcolor{yellow!15}0.95 & \cellcolor{yellow!15}0.80 
& 0.273&\cellcolor{yellow!15}0.670
& 0.273&\cellcolor{yellow!15}\underline{0.620} 
&0.517 &0.263 & \cellcolor{yellow!15}\textbf{0.600} \\

\cellcolor{blue!15}Industrial\&and\&Scientific & \cellcolor{blue!15}1.57 & \cellcolor{blue!15}1.18 & \cellcolor{blue!15}1.4 
&0.325 &\cellcolor{blue!15}0.830
&0.325 & \cellcolor{blue!15}\underline{0.826} 
& 0.508 & 0.314&\cellcolor{blue!15}\textbf{0.817} \\

\cellcolor{pink!15}Musical\&Instruments & \cellcolor{pink!15}1.34 & \cellcolor{pink!15}1.19 & \cellcolor{pink!15}1.17 
& 0.309&\cellcolor{pink!15}0.860 
& 0.309&\cellcolor{pink!15}\underline{0.850}
&0.610 &0.304 & \cellcolor{pink!15}\textbf{0.831} \\

\cellcolor{gray!15}Software & \cellcolor{gray!15}1.64 & \cellcolor{gray!15}1.74 & \cellcolor{gray!15}1.47 
&0.516 &\cellcolor{gray!15}1.350
&0.516&\cellcolor{gray!15}\underline{1.206}
& 0.7899& 0.505&\cellcolor{gray!15}\textbf{1.180} \\

\cellcolor{purple!15}Video\&Games & \cellcolor{purple!15}1.65 & \cellcolor{purple!15}1.46 & \cellcolor{purple!15}1.43
&0.355 & \cellcolor{purple!15}1.180
& 0.355&\cellcolor{purple!15}\underline{0.949}
&0.820  & 0.350&\cellcolor{purple!15}\textbf{0.939} \\
\bottomrule
\end{tabular}
}
}
\begin{tablenotes}
      \centering
      \item The results under the same color block are comparable, and best/runner-up results are highlighted in bold/underlined. 
    \end{tablenotes}
\label{main_exp}
\end{table*}

\subsection{Experimental Setup}
\subsubsection{Evaluation Metrics} Following~\cite{RGCL,LoID}, the overall performance is evaluated by \textbf{Mean Squared Error (MSE)}, a widely used metric for rating prediction in recommendation systems. 
However, for the CoT Generation stage, we specifically chose \textbf{Mean Absolute Error (MAE)} as the evaluation metric. 
This decision was made because MAE provides a more direct and interpretable measure of prediction deviation, offering a clearer indication of model performance without being disproportionately influenced by the dataset's imbalanced distribution. 
For review-based recommendation, a relative improvement exceeding 1\% is considered significant~\cite{evaluation}.

\subsubsection{Models for Comparison}
To comprehensively evaluate our approach, we compare the performance of RecCoT against several representative baselines.
The models were chosen to enable a thorough comparison of various recommendation methodologies.
The baselines include:

\begin{itemize}
    \item \textbf{ANR~\cite{anr}}: A model that leverages review text to extract multiple semantic dimensions, creating more nuanced user and item representations.
    \item \textbf{BiGI~\cite{Bigi}}: A framework that captures both the local community structure and long-range dependencies in a bipartite graph by maximizing mutual information.
    \item \textbf{RGCL~\cite{RGCL}}: A graph contrastive learning framework that uses a dual-task objective to align user-item interaction features with review semantics.
    \item \textbf{LoID~\cite{LoID}}: A cross-domain recommendation framework that utilizes a large language model and LoRA to achieve efficient, training-free knowledge transfer.
\end{itemize}

\subsubsection{RecCoT Model Variants}
To isolate the contributions of our key components, we evaluate two variants of our framework:

\begin{itemize}
    \item \textbf{LoID + Cache Embedding}: This model is an extension of the original \textbf{LoID} framework, augmented with our proposed cache mechanism. 
    This variant serves as a crucial point of comparison to specifically evaluate the effectiveness and impact of the cache mechanism on model performance.
    \item \textbf{RecCoT}: This is our most comprehensive variant. It combines information from user reviews and the explicitly generated Chain-of-Thought (CoT) information.
    This allows the model to leverage the structured reasoning steps extracted from reviews, providing a deeper understanding of user preferences and leading to more explainable and accurate recommendations.
\end{itemize}

\subsubsection{Implementation Details}
Our model's hyperparameters were meticulously set to ensure optimal performance and training stability. 
The embedding size, $d$, was fixed at \textbf{768}, with a consistent dropout rate of \textbf{0.5} applied across the entire model to prevent overfitting.
For the CoT Generation stage, we utilized the Qwen model with a batch size of \textbf{4} and a learning rate of \textbf{2e-4}.
In the Semantic Embedding Alignment stage, which employs BERT, the hyperparameters were uniformly set to an embedding size of \textbf{768}, a dropout rate of \textbf{0.5}, a learning rate of \textbf{1e-5}, and a batch size of \textbf{4}, trained for \textbf{4 epochs}. 
For the Down-streaming RecSys stage, the attention layer numbers, learning rate, batch size, and epochs were fixed at \textbf{4}, \textbf{1e-4}, \textbf{128}, and \textbf{5}, respectively. 
All model parameters were optimized using the Adam algorithm~\cite{adam}, and the entire model was implemented with Pytorch~\footnote{\url{https://pytorch.org}} based on an Nvidia RTX GPU.

\subsection{Performance Comparison and Analysis}

The main results for the rating prediction task on seven Amazon datasets are summarized in Table~\ref{main_exp}. 
The evaluation metric is MSE, where lower values indicate better performance. 
In the table, emb denotes the MSE from the model's pre-training stage, and CoT denotes the MAE from the CoT Generation stage.
From the table, we draw the following key observations:

\begin{itemize}
    \item \textbf{Overall Performance.} Our proposed model, \textbf{RecCoT}, consistently and significantly outperforms all baseline methods across all datasets, establishing a new state-of-the-art. 
    For instance, on the \textit{Amazon\&Fashion} dataset, RecCoT achieves an MSE of \textbf{0.570}. 
    This represents a substantial improvement over the strongest graph-based baseline, RGCL (1.44), and a notable 16.2\% relative improvement over the powerful LLM-based LoID (0.680).

    \item \textbf{Effectiveness of Cache Embedding.} The ablation results clearly demonstrate the efficacy of our proposed cache mechanism. 
    The \textbf{LoID + Cache Embedding} variant consistently surpasses the original LoID model, achieving the runner-up performance on all datasets. 
    On the \textit{Video\&Games} dataset, the introduction of the cache reduces the MSE from 1.180 to \textbf{0.949} (a 19.6\% relative improvement), highlighting that providing the model with relevant historical context via the cache is a highly effective strategy.

    \item \textbf{Contribution of Chain-of-Thought (CoT).} The full \textbf{RecCoT} model further improves upon the already strong performance of the cache-enhanced variant. 
    This demonstrates the value of incorporating explicit reasoning. For example, on the \textit{Software} dataset, RecCoT (1.180) improves upon LoID + Cache Embedding (1.206). 
    This suggests that by explicitly modeling the user's reasoning process through CoT, our model can capture more nuanced preferences, leading to more precise and explainable recommendations.
\end{itemize}

Our experiments validate that both the cache mechanism and the Chain-of-Thought component are crucial for the model's success. 
Their combination allows RecCoT to effectively leverage review information and achieve superior recommendation performance.
\subsection{The Effectiveness of Reward Mechanism}
\begin{table}[t]
\small
\centering
\caption{Effectiveness of Proposed Reward Mechanism: An Ablation Study}
\resizebox{\linewidth}{!}{
\begin{tabular}{lccc}
\toprule
\textbf{Reward Mechanism} & \textbf{CoT(MAE)} & \textbf{Emb} & \textbf{RecSys} \\
\midrule
\multicolumn{4}{c}{\textbf{Amazon Fashion}} \\
\cdashlinelr{1-4}
Linear Reward & 1.1 & 0.421 & 0.789 \\
Frequency Reward & 0.8 & 0.401 & 0.570 \\
\midrule
\multicolumn{4}{c}{\textbf{Musical Instruments}} \\
\cdashlinelr{1-4}
Linear Reward & 0.66 & 0.314 & 0.854 \\
Frequency Reward & 0.61 & 0.304 & 0.831 \\
\bottomrule
\end{tabular}
}
\label{tab:Reward}
\end{table}

In the Amozon datasets, the label distribution is heavily skewed towards positive ratings, with a rating of 5 accounting for over 80\% of the data.
This typical long-tailed distribution can bias models toward high-frequency samples, consequently impairing their predictive performance on less frequent, or "long-tail" items.

To mitigate this issue, we introduce a novel reward mechanism within our RecCoT framework, specifically designed to counter the model's inherent preference for dominant classes.
To demonstrate the efficacy of our proposed mechanism, we conducted an ablation study, comparing it against a standard linear reward strategy. 
As illustrated in Table~\ref{tab:Reward}, we evaluated the models across three key stages: CoT(CoT Generation), Semantic(Semantic Embedding Alignment), and Downstream(Down-streaming RecSys).
%

The experimental results unequivocally show that our designed reward strategy significantly outperforms the linear alternative. 
This finding provides strong evidence for the effectiveness of our mechanism in addressing the challenges posed by long-tailed data distributions.

\subsection{The Effectiveness of User Review Count}

\begin{table}[tbp]
\footnotesize
\centering
\caption{Review Count Performance comparison}
\resizebox{\linewidth}{!}{
\begin{tabular}{lcccc}
\toprule
\textbf{User Group} & \textbf{Percent} & \textbf{LoID+ Cache Embedding} & \textbf{RecCoT} \\
\midrule
\multicolumn{4}{c}{\textbf{Amazon Fashion}}         \\
\cdashlinelr{1-4}
Low(<10) &86.47\% &0.633	&	0.570\\
Mid(<20)& 10.86\% &0.765		&	0.718\\
High(>20)& 2.67\% &0.776		&	0.702\\
\midrule
\multicolumn{4}{c}{\textbf{Musical Instruments}}    \\
\cdashlinelr{1-4}
Low (<10) & 72.47\% &0.892		&	0.831\\
Mid (<20) & 20.97\% &0.922		&	0.83\\
High (>20) & 6.56\% &0.903	&	0.79\\
\bottomrule
\end{tabular}}
\label{tab:user length}
\end{table}
In addition to our reward mechanism, we designed a cache mechanism to store and leverage a user's historical interaction information. 
To quantify the impact of this history on model performance, we conducted a comparative analysis of model performance across different user engagement levels, with user groups categorized by their review count. 
This study was performed on two distinct datasets: Amazon Fashion and Musical Instruments, with the user distribution for each detailed in Table~\ref{tab:user length}.

The results in Table~\ref{tab:user length} show a consistent performance trend for both datasets. 
For the Amazon Fashion dataset, which is dominated by low-engagement users (86.47\%), the RecCoT model achieved its highest score of 0.570 for the Low user group (<10 reviews).
Performance declined for the Mid (<20 reviews) and High (>20 reviews) user groups, with scores of 0.718 and 0.702, respectively.

A similar pattern is observed for the Musical Instruments dataset, which has a higher average number of interactions. 
Performance for the RecCoT model was highest for the Low user group (<10 reviews) at 0.831 and gradually decreased for the Mid (<20 reviews) and High (>20 reviews) user groups, with scores of 0.830 and 0.790.

These findings collectively demonstrate that a one-size-fits-all approach to leveraging user history is suboptimal. 
The results for both datasets indicate that while user history is a valuable signal, a greater volume of history (i.e., from users with a high review count) does not necessarily lead to improved performance, and may even be a source of noise. 
The results underscore the necessity of a mechanism, such as our proposed cache mechanism, that can effectively adapt to varying historical contexts and user demographics.

\subsection{Discussion of Semantic Generalization}
\begin{table}[th]
\centering
\caption{Performance Evaluation of RecCoT in Semantic Generalization}
\resizebox{\linewidth}{!}{\begin{tabular}{lccc}
\toprule
\textbf{Target} & \textbf{CoT\&Emb ckpt Source} & \textbf{RecSys} \\
\midrule
\multirow{2}{*}{Amazon Fashion} & Self & 0.570 \\
& CDs and Vinyl &0.580 \\
\cdashlinelr{1-3}
\multirow{2}{*}{Industrial and Scientific} & Self & 0.824 \\
& CDs and Vinyl &0.850 \\
\cdashlinelr{1-3}
\multirow{2}{*}{Musical Instruments} & Self & 0.831 \\
& CDs and vinly &0.840 \\
\bottomrule
\end{tabular}}
\label{tab:Generalization}
\end{table}
We investigate the semantic generalization and transferability of the RecCoT framework. 
Specifically, we assess its generalization ability by applying the model, which was pre-trained on the CDs and Vinyl product category, directly to three distinct new product areas—Amazon Fashion, Industrial and Scientific, and Musical Instruments—without any fine-tuning.
The goal is to evaluate the model's ability to transfer knowledge learned from the source category to downstream tasks.

As shown in Table~\ref{tab:Generalization}, the model demonstrates robust performance across all three target categories. 
The performance of the transferred model is highly competitive with the category-specific benchmarks. For instance, in the Amazon Fashion category, the transferred performance is negligibly different from the benchmark for that specific product field.
Similarly, in the Industrial and Scientific and Musical Instruments fields, the results remain remarkably close to their respective baselines.

These findings collectively validate the strong generalization and transferability characteristics of our model. 
The ability to transfer knowledge from a source category to a new product area and achieve performance that is highly competitive with a category-specific benchmark, all without any retraining, is a testament to the model's effectiveness and robustness. 
This indicates that the model has learned generalizable semantic representations rather than patterns specific to one product field.

\subsection{Discussion of User Review Length}
\begin{table}[t]
\centering
\caption{Impact of User Review Length on Model Performance (Musical Instruments Dataset)}
\setlength{\tabcolsep}{8pt}{
\begin{tabular}{lcc}
\toprule
\textbf{Review Length Range (Characters)} & \textbf{MSE} \\
\midrule
0-127 & 0.2825 \\
128-255 & 0.2464 \\
256-511 & 0.2297 \\
512-1024 & 0.3501 \\
\bottomrule
\end{tabular}
}
\label{tab:review_length}
\end{table}

Fusing our generated Chain-of-Thought (CoT) with original reviews presents a challenge by significantly increasing text length, which can impact the ability of LLMs to process information effectively. 
To investigate this relationship, we conducted a systematic ablation study on the Musical Instruments dataset to understand the effect of review length on model performance.
The results, detailed in Table~\ref{tab:review_length}, reveal a clear non-monotonic trend: model performance doesn't simply improve with more text but instead peaks and then declines after a specific threshold.

Specifically, the model achieves its peak performance with moderately sized reviews, where it acquires a rich, yet focused, set of information for effective reasoning. 
Conversely, excessively long reviews (e.g., over 512 characters) introduce a high degree of information redundancy and semantic noise, which can interfere with the model's ability to extract key insights. 
This information overload can cause the model to suffer from attention drift, causing a performance drop. 
Therefore, our CoT fusion strategy is specifically designed to generate text that contains critical reasoning information while carefully avoiding unnecessary length, thereby maximizing the model's benefit from the reviews without introducing detrimental noise.

\subsection{Discussion on the Explainability of CoT}
The use of CoT provides a transparent mechanism for understanding our model's rating predictions, allowing us to trace the reasoning from the raw review text to the final score. 
The provided cases demonstrate that while the CoT approach is generally effective at capturing core sentiment and guiding predictions, it also presents valuable insights into the model's limitations and behavior.

\subsubsection{Strengths in Capturing Sentiment and Nuance}
The CoT mechanism is highparticularly effective atin handling reviews with a clear or mixed sentiment, providing a transparent and accurate basis for our model's predictions. 
sentiment or a mix of positive and negative points. 
For straightforward negative reviews, as seen in \textbf{Case 1} and \textbf{Case 7}, the generated CoT accurately identifies the strong criticisms---such as ("common strap," or "crap waste of money"---), leading to predictions (1.8) that closely align with the low user ratings (1.0). 
%
%
Similarly, for multi-faceted reviews, like in (\textbf{Case 2}), the CoT successfully captures the dual sentiments ("price is good," but "quality is so so"). 
This and guides the model to a balanced prediction (3.7) that accurately reflects the mixed nature of the feedback. 

\subsubsection{Limitations and Diagnostic Insights}
The true value of the CoT is often revealed in cases where predictions are misaligned or where the CoT itself contains flaws, as these instances serve as a powerful diagnostic tool for model improvement.
\begin{itemize}
    \item \textbf{Misinterpretation and Incompleteness:} The CoT generation process is not infallible. 
    In \textbf{Case 4}, a clearly positive review is misclassified as having "moderate" sentiment due to non-existent criticisms, leading to a lower-than-expected prediction (3.8 for a 4.0 rating). 
    This highlights a failure in the CoT to correctly interpret the overall tone. Similarly, in \textbf{Case 3}, the CoT is accurate but incomplete, failing to capture the reviewer's strong analogy ("looks designed more for a tennis racket"). 
    This, which may have contributed to the model's high prediction (2.5) for a very negative review (1.0).

    \item \textbf{Robustness against Noisy Labels:} The CoT mechanism demonstrates a preference for textual evidence over potentially noisy numerical ratings. 
    In \textbf{Case 5}, the review text is intensely negative ("cracked and exploded"), yet the user's rating is a contradictory 5.0. 
    The CoT correctly summarizes thise negative feedback, and the model, guided by this textual reasoning, produces a moderate prediction (3.8). 
    This suggests that the model is robust against misleading true labels by grounding its reasoning in the detailed review content.

    \item \textbf{Difficulty with Technical Jargon:} \textbf{Case 6} is particularly insightful for identifying model weaknesses. The review uses technical, positive language to describe a product's benefits ("increased low end, punch and power"). 
    However, the CoT incorrectly labels the sentiment as "neutral," resulting in a poor prediction (3.5) that is far from the user's 5.0 rating. 
    This failure case is invaluable, as the flawed CoT explicitly reveals a specific model weakness---: its difficulty in interpreting domain-specific technical jargon---, providing a clear path for future improvements.
\end{itemize}

\section{Conclusions}
In this paper, we introduce RecCoT, a simple yet effective two-stage framework that addresses the challenge of generating interpretable and accurate recommendations. 
Our approach balances the high-quality reasoning of a large language model (LLM) with the fast, efficient demands of a recommender system. 
The framework consists of two stages: first, a large LLM generates high-quality Chain-of-Thought (CoT) explanations using a novel reward-constrained reasoning paradigm; second, a smaller LLM (BERT) computes fast semantic embeddings that are aligned via contrastive learning.
Extensive experiments on benchmark datasets show that RecCoT consistently surpasses state-of-the-art methods, and in-depth analyses further underscore the effectiveness of our model's individual components.
We will explore more advanced policies to prevent reward hacking in future work.

%

\section{CoT Case}
\begin{codebox}[Case: 1]
\textbf{review:} for the money its just a common strap you would see on any duffle bag. looks nothing like a guitar strap..way too thin as well.

\textbf{CoT:} This review is negative and critical of the product's material quality and price point. It highlights that it's not a great value for money and compares it unfavorably to other similar products. The reviewer seems disappointed with the product.

\textbf{user rating:} 1.0 \space\space\space\space\space\space\space\space\space\space\space\space\space\space\space\space\space\space\space\space     
\textbf{predict:} 1.8
\end{codebox}

\begin{codebox}
[Case: 2]
\textbf{review:} price is good, quality is so so. sizing is little too small. 

\textbf{CoT:} This product has received mixed reviews based on the given user feedback. The customer expresses that the price is good but the quality seems to be subpar. Additionally, they mention that the sizing is slightly too small. While the initial impression might not be overwhelmingly negative due to the good price, the minor criticism of the sizing suggests some dissatisfaction.

\textbf{user rating:} 3.0
\space\space\space\space\space\space\space\space\space\space\space\space\space\space\space\space\space\space\space\space     
\textbf{predict:} 3.7
\end{codebox}

\begin{codebox}[Case: 3]
\textbf{review:} not even one storage pocket, and zipper is in bad location. the neck sleeve is too tight for stock classical tuning keys. looks designed more for a tennis racket.

\textbf{CoT:} The user has provided some negative feedback about the product's design, particularly regarding the storage pockets and neck sleeve. They mention that the key looks designed more for a tennis racket and suggest it may not be suitable for their needs.

\textbf{user rating:} 1.0
\space\space\space\space\space\space\space\space\space\space\space\space\space\space\space\space\space\space\space\space     
\textbf{predict:} 2.5
\end{codebox}

\begin{codebox}[Case: 4]
\textbf{review:} works well and great for the price.

\textbf{CoT:} This is an interesting review about a product that works well and is good value for money. The reviewer describes it as a great fit for the budget, which is quite positive. However, the overall sentiment seems to be moderate due to some minor criticisms.

\textbf{user rating:} 4.0
\space\space\space\space\space\space\space\space\space\space\space\space\space\space\space\space\space\space\space\space     
\textbf{predict:} 3.8
\end{codebox}

\begin{codebox}[Case: 5]
\textbf{review:} this is not a good buy. within days of stringing to normal tension, the bridge, which is thinly designed and set at an angle asking to be broken ... well the bridge cracked and exploded off the instrument. in a panic because beyond that the instrument did as i had hoped, i promptly ordered another one. a repeat history, within hours of being strung to normal tension, the bridge was already suffering a fracture. im a professional musician  depend on my instruments to performquite unnaceptable, never before experienced failure with an ukulele.

\textbf{CoT:} This review expresses strong dissatisfaction with the product, describing it as a 'bad buy' due to several issues including cracking, explosion, and structural failures. The reviewer mentions having previously relied on their instrument for performance, indicating they have a high level of trust in the product.

\textbf{user rating:} 5.0
\space\space\space\space\space\space\space\space\space\space\space\space\space\space\space\space\space\space\space\space     
\textbf{predict:} 3.8
\end{codebox}

\begin{codebox}[Case: 6]
\textbf{review:} a single ply of 12mil film that delivers increased low end, punch and power when compared to traditional 10mil single ply heads. ideal for heavier hitters looking for a solid sound and more life to the head but prefer the feel of a singleply. 

\textbf{CoT:} This review seems to be neutral towards the product's performance. The reviewer mentions that it provides increased low-end, punch, and power compared to traditional 10mil single ply heads. They also recommend it as ideal for heavier hitters who value a solid sound and more life to their head. However, they do not seem to have any specific positive or negative sentiments.

\textbf{user rating:} 5.0
\space\space\space\space\space\space\space\space\space\space\space\space\space\space\space\space\space\space\space\space     
\textbf{predict:} 3.5
\end{codebox}

\begin{codebox}[Case: 7]
\textbf{review:} crap waste of money.

\textbf{CoT:} This is a negative review mentioning that the product is worthless and has poor value for money. The user seems disappointed with their experience and does not seem to have any positive aspects to share.

\textbf{user rating:} 1.0
\space\space\space\space\space\space\space\space\space\space\space\space\space\space\space\space\space\space\space\space     
\textbf{predict:} 1.8
\end{codebox}

\begin{codebox}[Case: 8]
\textbf{review:} needed some nl cables for an install. these fit the bill. good build quality. cable seems to be of good quality. note these are only 2 conductor cables with nl4 connectors. these will work on passive speakers, but not biamped speakers. would recommend.

\textbf{CoT:} This review provides some basic information about the product's features and user satisfaction. The reviewer mentions that the cables need some NL cables for an installation and that they seem to meet the needs of passive speakers. However, it also notes that the cables may not be suitable for biamped speakers due to their limited number of conductors.

\textbf{user rating:} 5.0
\space\space\space\space\space\space\space\space\space\space\space\space\space\space\space\space\space\space\space\space     
\textbf{predict:} 3.8
\end{codebox}

\balance
\bibliographystyle{ACM-Reference-Format}
\bibliography{sample-base-extend.bib}
\end{document}